# Exciton Liquid in Coupled Quantum Wells

Michael Stern[1]*†, Vladimir Umansky[1], Israel Bar-Joseph[1]

[1]Department of Condensed Matter Physics, The Weizmann Institute of Science, Rehovot, Israel.

*Corresponding author Email: michael.stern@cea.fr.

†Current Adress: Quantronics Group, SPEC, IRAMIS, DSM, CEA Saclay, Gif-sur-Yvette, France.

Excitons in semiconductors may form correlated phases at low temperatures. We report the observation of an exciton liquid in GaAs/AlGaAs coupled quantum wells. Above a critical density and below a critical temperature the photogenerated electrons and holes separate into two phases, an electron-hole plasma and an exciton liquid, with a clear sharp boundary between them. The two phases are characterized by distinct photoluminescence spectra and by different electrical conductance. The liquid phase is formed by the repulsive interaction between the dipolar excitons, and exhibits a short range order, which is manifested in the photoluminescence lineshape.

More than forty years ago it was shown theoretically that excitons in semiconductors may form a condensate at cryogenic temperatures (*1*). Subsequently, a wealth of classical and quantum phases, ranging from electron-hole plasma to superfluid liquid, have been predicted to occur in this material system (*2-4*). Spatially indirect dipolar excitons with the electron and holes residing in separate coupled quantum wells (CQW) offer a promising test bed for these phenomena (*5-8*). They are characterized by a strong repulsive dipole-dipole interaction, which can be easily engineered by designing the sample structure. Their density, $n$, and the distance between the centers of the wells, $d$, are key parameters that govern the formation of condensed states in this system. Together they set the strength of the dipole interaction with the surrounding carriers. The interaction energy, $\delta E(n) = \iint V(r) n(r) d^2 r$, can be expressed in terms of an effective radius, $r_0$, as

$$\delta E(n) = \frac{\langle n \rangle e^2}{\varepsilon} \left[ \sqrt{r_0^2 + d^2} - r_0 \right]$$

Here $r_0$ is a characteristic radius around every exciton within which the carrier density is depleted as a consequence of the repulsive interaction $V$, $\langle n \rangle$ is the average density, e the electron charge and $\varepsilon$ is the dielectric constant. $\delta E(n)$ can be directly measured in a photoluminescence experiment as an energy shift of the recombination energy with density.

Theory predicts that at densities well above the Mott transition the system may form an exciton liquid (*3-4*). For $d$ smaller than the exciton Bohr radius $a_B$ the exchange interaction can compensate the dipole repulsion and a quantum liquid is formed. For $d \geq a_B$ this phase is not stable, but the system may form a classical liquid. The origin of this classical liquid is related to the formation of a region around each exciton, with radius $\rho \approx 4d^2/a_B$, where the wavefunctions of other excitons are exponentially small (*4*). When $n\rho^2 \geq 1$, the wave functions of neighboring excitons do not overlap, and quantum correlations are suppressed. Classical correlations, on the other hand, are very strong and may give rise to short range order. This classical limit of strongly interacting dipoles is the focus of our work.

The system we study is formed by two GaAs quantum wells with well widths of 12 and 18 nm separated by a 3 nm $Al_{0.28}GaAs$ barrier (energy diagram shown in Fig. 1A); the distance between the centers of the wells is thus $d = 18$ nm, almost twice as large as $a_B$. The carriers are photo-generated by a laser and separated by the application of a gate voltage $V_g$ in the direction perpendicular to the wells. More experimental details are given in (*9*).

In Fig. 1B we show the evolution of the spectrum with increasing laser power *P*. Between 1-100 µW the indirect recombination energy (*IX*) increases monotonically with *P* as the density *n* increases. This shift to higher energy is the manifestation of the growing interaction energy, $\delta E(n)$, and it allows us to determine the e-h density: for example, at 20 µW $\delta E = 9$ meV, which corresponds to $\sim 2-3 \times 10^{10}$ cm$^{-2}$. $\delta E$ is not linear with power, and saturates as the *IX* peak approaches the wide-well exciton line, $X_{WW}$. This behavior can be understood noting that the recombination rate Γ is proportional to the overlap integral of the electron and hole wavefunctions (*10*): as the density increases, the potential drop between the quantum wells is screened, and Γ increases exponentially. An interesting implication of this behavior is that at *P* >>20 µW the density across the sample is almost constant, and depends only weakly on the local power.

In Fig. 1C, we show the spectra at various powers, where the blueshift of the *IX* peak with power is clearly evident. We notice that the *IX* spectra at all powers are characterized by a pronounced exponential tail at the low energy side (*11*). This lineshape reflects the broad range of interaction energies (and corresponding $r_0$ values) that can be realized in an e-h plasma in CQW. We find that the exponential tails of the lines at low and medium powers merge with the high power spectra. Indeed, one can reproduce the essential features of the observed behavior by calculating the interaction of a recombining dipole with the surrounding gas, taking into account the change of Γ with energy (Fig. 1D). Hence, this low energy tail can be viewed as the spectral signature of interacting e-h plasma in CQW. More details are given in Sec. 1.1 of (*9*).

Let us turn now to the behavior at higher power. At *P*=210 µW a new line, denoted by *Z*, appears below the *IX* line, and shifts only slightly with power. One can show that this line is also due to indirect recombination of electrons in the narrow well with holes in the wide well: it shifts linearly with the gate voltage, approximately parallel to the *IX* line, remaining 3-4 meV above $E_{IX}^0$, the *IX* energy at low power, <1 µW.

We find that the Z line comes from a well defined spatial region in the illuminated area. We image the photoluminescence from the mesa and observed that below threshold the image is smooth and follows the Gaussian profile of the excitation spot. Above threshold a striking behavior appears: the luminescence separates into two regions, with a narrow (resolution limited) dark border between them (Fig. 2A). The location of this boundary is unrelated to any structural parameter of the sample, and can be affected by slight shifts of the exciting laser spot (Movies S1-S2).

The spectral properties of the two regions are analyzed by local photoluminescence measurements, with a 5 µm spatial resolution (Fig. 2B). We find that as we move across the boundary the spectrum changes from being dominated by *IX* in the region denoted in the figure as I to being dominated by *Z* in Region II.

In the following we show that the *Z* line is a manifestation of an exciton liquid:

(i) Criticality: We find that the transition of the system into the phase-separated state is thermodynamic and occurs below a critical temperature and above a threshold power. Figure 2C shows the relative area of Region II as a function of temperature. It is seen that this phase appears abruptly at T=4.7 K. Figure 2D demonstrates the power dependence: the appearance of Region II above a threshold power is clearly seen. We note that, in this case, the area of Region II extends gradually until it covers the entire sample.

(ii) High density: Time resolved measurements reveal that the lifetime of the carriers in Region II is ~60-80 ns (Fig. 2E), approximately twice as large compared to Region I. The phase separation therefore manifests two density regions: Region I with lower density and Region II with higher density. Using the absorption value of the CQW (12) we can estimate the density in Region II to be $\sim 4-6\times 10^{10}$ cm$^{-2}$ (9). At this high density we get $n\rho^2 \approx 10$ fulfilling the condition for a classical liquid, $n\rho^2 >> 1$. We note that the Z line energy is shifted relative to $E_{IX}^0$ by $\delta E = 3-4$ meV only. Using Eq. (1) we find that this corresponds to $r_0 \approx 25-35$ nm, approximately half the average interparticle distance at these densities.

(iii) Low diffusivity: Figure 3A shows the local intensity of the *IX* line as a function of the local excitation power. It is clearly seen that the *IX* intensity is well correlated with the excitation power in Region II, and is uncorrelated with it in Region I. This implies that in Region I the carriers can recombine away from where they were excited, while in Region II they are localized. The high diffusivity of the carriers in Region I is also seen when we compare the luminescence intensity profile in this Region before and after the phase separation (Fig. 3B). It is seen that the carriers in Region I are pushed away from the region boundary and aggregate at the sides of the mesa.

The localization of carriers in Region II is corroborated by in-plane electrical transport measurement conducted in a 4-probe geometry (Fig. 3C). Figure 3D shows the conductance of the sample, σ, as a function of power. The onset of phase separation is clearly manifested by a drop of the conductance value and a reduction of the slope of the curve. Above the threshold power, the sample is divided into two regions, each with a different conductance, with $\sigma_I >> \sigma_{II}$ (see the Materials and Methods section). The existence of a high density region with low conductance is consistent with an exciton liquid.

(iv) Ordering: An important insight comes from studying the lineshape of the *Z* line (Fig. 4A). It is seen that the characteristic low energy tail of the e-h plasma is absent and the line is an almost perfect Gaussian (Fig. 4B). Noting that the low energy tail is a manifestation of multiple values of $r_0$, which characterize the gas phase, we are led to conclude that the Gaussian shape represents a narrow distribution around a dominant configuration. We used a local photoluminescence measurement to extract the *Z* line at various locations and study its linewidth dependence on density. Remarkably, we find that the width of this Gaussian line decreases with increasing density (Fig. 4C). This line narrowing cannot be explained by an uncorrelated gas, where the linewidth should grow with density (*13*). To explain this behavior we conduct a simple Monte Carlo calculation, assuming a fixed depletion region of radius $r_0$ around each exciton. With this assumption the calculation reduces to inserting randomly N dipoles on a grid with $r_0$ periodicity, and calculating their interaction energy. We find that the linewidth is indeed Gaussian, and decreases with increasing density when $n\pi r_0^2 \geq 1/2$. Above this density the number of possible configurations starts to decrease, giving rise to a narrower range of interaction energies. In fact, by taking $r_0 \approx 25$ nm, the value obtained from the estimated density, we get a quantitative agreement with the observed behavior (Fig. 4C).

The accumulation of these findings indicates that a strongly correlated ordered phase appears below a critical temperature and above a threshold density. The sample separates into two phases: an interacting e-h plasma in Region I and a liquid phase in Region II. We note that Region II should not be perceived as a continuous liquid phase. The fact that we observe both *Z* and *IX* lines in Region II (Fig. 2B) indicates that this region consists of a mixture of droplets and gas. Indeed, the existence of such a mixture was predicted by calculating the free energy of an exciton gas and solid (*14*).

Several other experimental findings are consistent with this interpretation. The first is the existence of a dark boundary between the two phases (Fig. 2A). This unusual behavior is explained by the fact that the carriers in the gas region experience a strong repulsive force from the localized high density dipoles at the boundary. This repulsion creates a depletion region at their interface, and induces a current of e-h pairs towards the sides of the mesa (Fig. 3B). Another supporting evidence is related to the phase separation dynamics. We find clear evidence for a nucleation process: There is a relatively long (seconds) delay time between the onset of the laser light and the appearance of the phase separation (Fig. S3). After this delay time Region II spreads from a corner of the mesa to its full extent (Movie S3).

In closing, we wish to point out an unexplained observation. By comparing the integrated photoluminescence intensity before and after the phase separation we find that the recombination in Region II is not entirely radiative, and approximately 15% of the intensity is lost. Furthermore, we find that this darkening disappears at a weak perpendicular magnetic field, $B \sim 0.3$ T. Above this field the phase separation is still observed. The fact that the phase separation can be observed without darkening rules out the formation of a dark condensate as the source of the observed behavior (*8,15*). Further studies are needed to examine this behavior.

We would like to thank R. Suris, B. Laikhtman, M. & R. Combescot, and V. Steinberg for fruitful discussions and M. Brook for the magnetic field measurements. This work was supported by the Israeli Science Foundation.

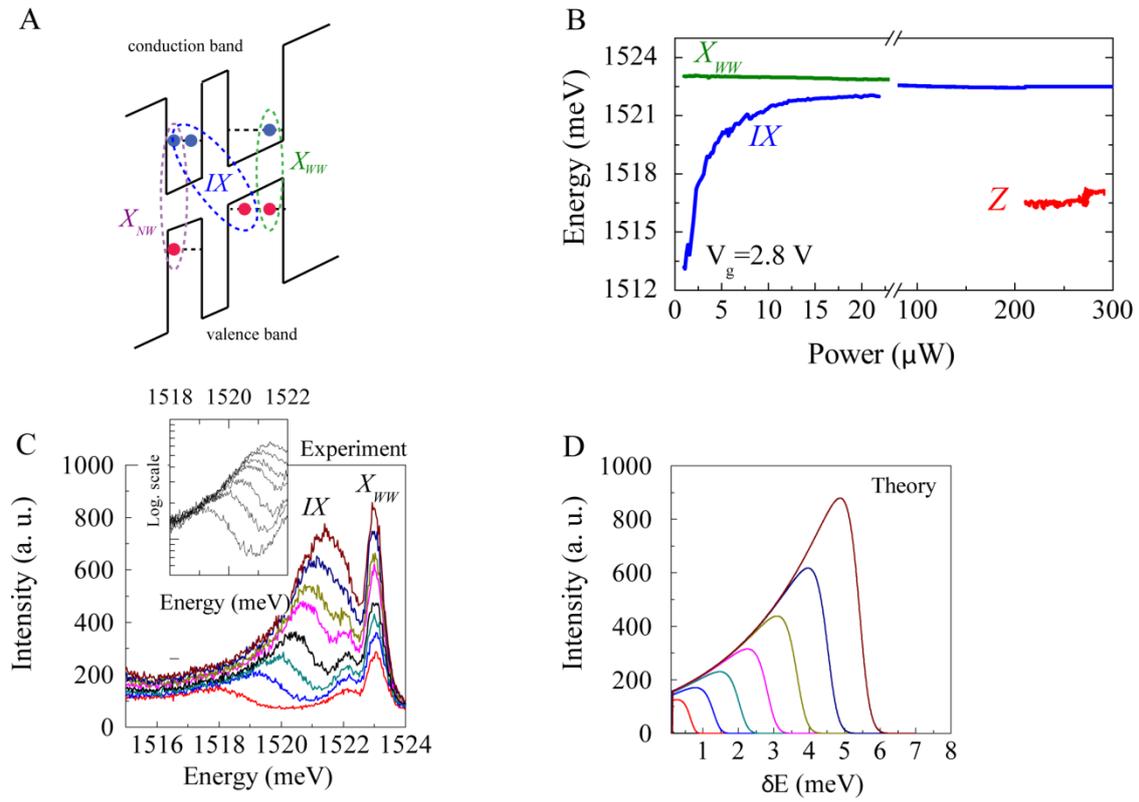

**Fig. 1**. **The evolution of the photoluminescence spectrum with power.** (**A**) The energy band diagram of the asymmetric CQW under perpendicular applied electric field. $X_{WW}$ ($X_{NW}$) is the exciton in the wide (narrow) well and *IX*- the inter-well indirect recombination. The blue (red) circles represent electrons (holes). (**B**) Photoluminescence peak positions for $X_{ww}$, *IX*, and *Z* lines as a function of laser power at T=1.5 K. (**C**) The evolution of the spectrum with power between 3 to 10 μW (The small peak at 1522 meV is the trion). The inset shows the spectra on a logarithmic scale. (**D**) Calculated spectra demonstrating the formation of the low energy exponential tail.

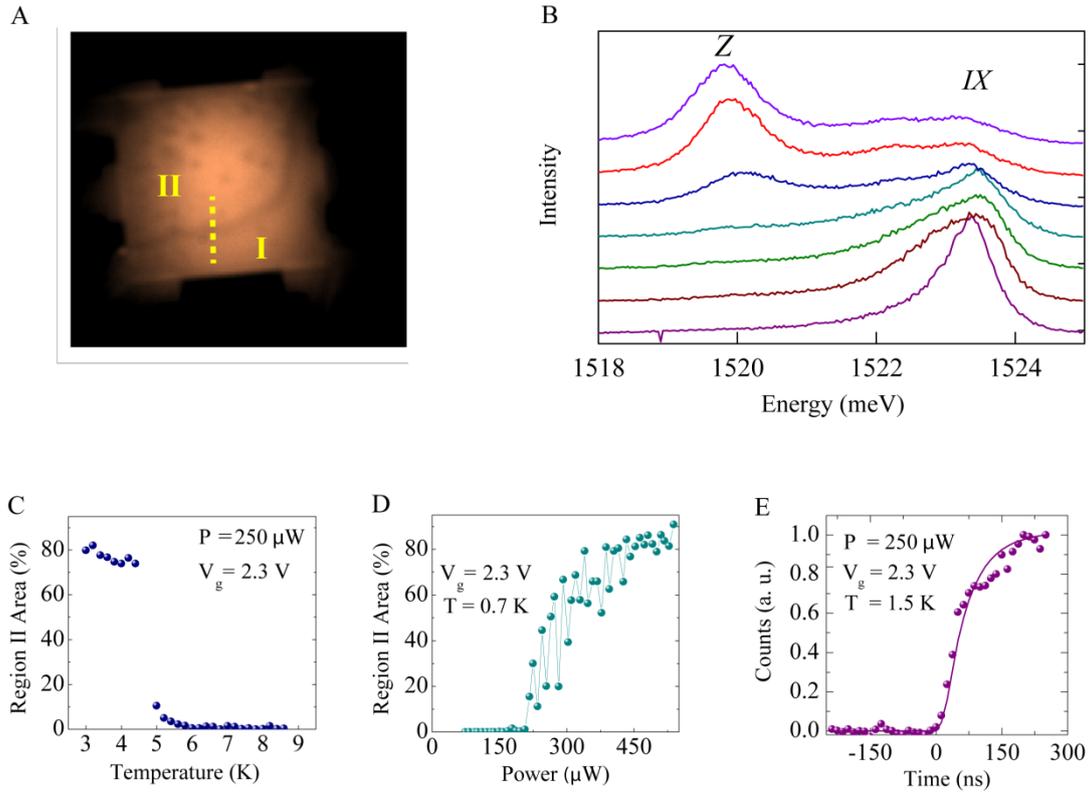

**Fig. 2. The phase separation.** (**A**) Image of the photoluminescence from the mesa showing the phase separation between Region I and Region II. The dark spots correspond to defects in the sample, which give reduced luminescence intensity in Region II. This measurement was performed at T=1.5 K, $V_g$=2.3 V and P=250 μW. (**B**) Spectra measured along the dotted yellow line in (**A**) in the vicinity of the phase boundary. Each spectrum is shifted vertically by a fixed value, corresponding to its location along the line. (**C-D**) Evolution of the area of Region II as a function of temperature (**C**) and laser power (**D**). The noise in the measurements reflects the fact that at some power levels, especially near threshold, the phase boundary exhibits large temporal fluctuations. (**E**) The rise of the Z line as a function of time measured by correlated photon counting using an acousto-optic modulator. At t=0, the laser is switched on and persists for ~ 4 μs. The solid line shows a fit to the experimental data which gives 70±10 ns lifetime.

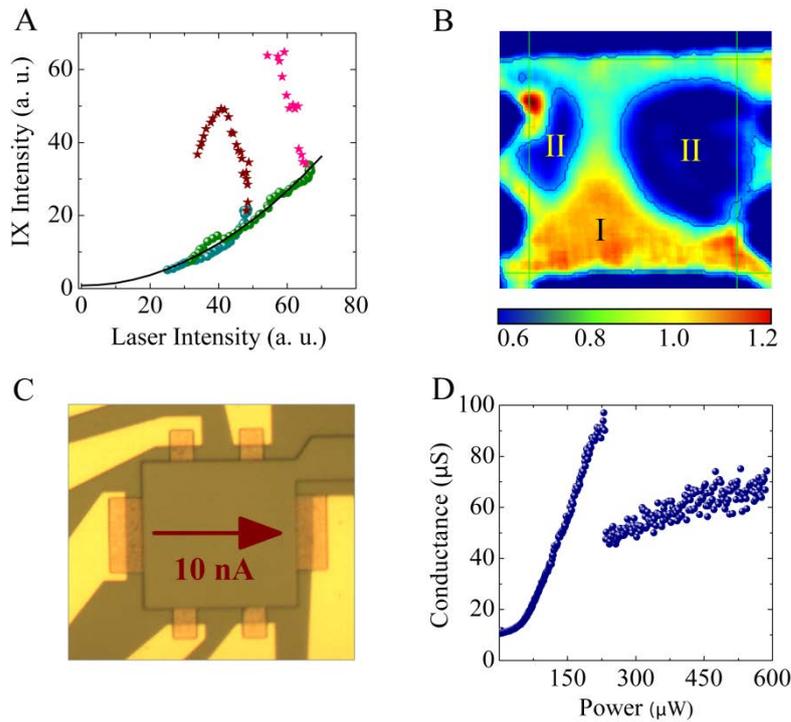

**Fig. 3. The low diffusivity of Region II.** (**A**) A measurement of the local *IX* intensity as a function of the local laser intensity along two lines, similar to the dotted line in Fig. 2A. The brown and pink stars are measured in Region I, while the blue and green dots in Region II. The parabolic solid line demonstrates the correlation between the *IX* and laser intensities in Region II. (**B**) The ratio between the photoluminescence images taken before and after phase separation. The thin green lines indicate the position of the mesa and are guide for the eyes. The blue color shows a reduction of intensity (<0.75) after phase separation while the red color represents an increase (>1.1). The diffusion away from the phase separation line towards the edges of the sample is clearly seen in Region I. (**C**) Optical microscope image of the sample showing the 4-probe electrical contacts to the 100 μm-square mesa. (**D**) 4-probe conductance as a function of laser power at T=1.5 K, $V_g$=2.3 V. The abruptness of the drop is due to the fact that the time interval between the measured points was one second, which is shorter than the nucleation time at 200 μW (*11*).

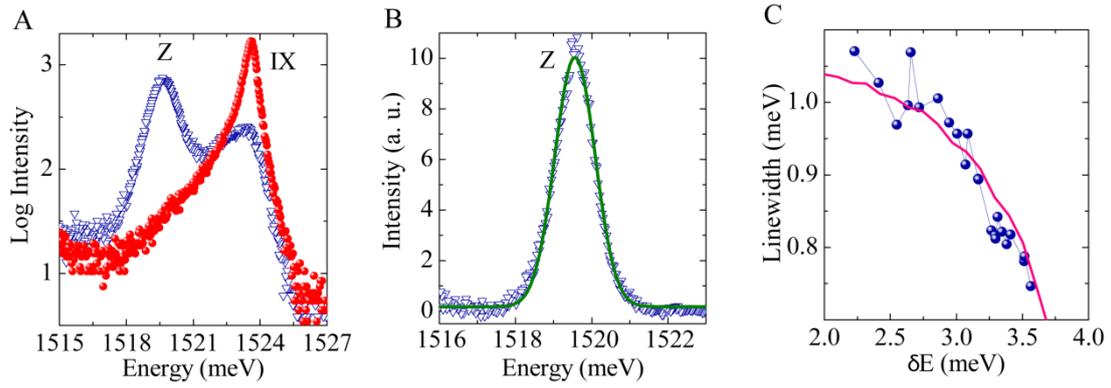

**Fig. 4. Shape and width of the Z line.** (**A**) A comparison between the photoluminescence spectrum in Region I (red) and Region II (blue). (**B**) The spectrum of the Z line obtained after removal of the exponential energy dependence, Γ(E). The solid line shows an excellent fit to a Gaussian. (**C**) The width of the Z line as a function of the energy shift $\delta E$. The reduction of the linewidth with density is clearly seen. The solid line shows the width obtained by Monte Carlo simulations taking into account 0.4 meV inhomogeneous broadening (HWHM). A good agreement between measurement and calculation is obtained for $r_0$=25 nm.